# Integral equation for the interfacial tension of liquid metal in contact with ionic melt

Oleg A. Kobelev* and Alexandr V. Kobelev**

*Urals State Technical University UGTU, Yekaterinburg 620002,*
**Inst of Metal Physics, Urals Branch of RAS, Yekaterinburg 620219,*
*Russian Federation*

Abstract

The closed integral equations for the interfacial tension as a function of external polarization at the liquid metal – ionic melt interface are derived. The version of Popel' – Pavlov isotherm is applied to the analysis of electrocapillary curves (ecc), i.e. the dependences of interfacial tension on electrode potential. The interaction between adsorbed particles is taken into account within the 'two exchange parameters' approximation. The type of the distribution of electric potential in the double electric layer (del) is assumed to be like 'in series connected capacitors'. The methods of solution are proposed for the analysis of the experimental ecc's.

## 1. Introduction

The adsorption of components under external electric polarization at the liquid metal - oxide melt interface has been studied, as a rule, on the basis of the electrocapillary equation and the ecc tracing back to the classical concept of the surface tension (see, for instance, [1-4]).

In the previous studies devoted to the ecc analysis [5-7], the model has been developed which takes into account the competitive adsorption of particles shifting a zero-charge potential in Langmuir's isotherm approximation. The model has allowed to describe quantitatively the electrocapillary features of the liquid metal - oxide melt interface. However, the reliability of the obtained data concerning processes at the investigated boundary is low, apparently owing to a number of specific assumptions that have to be made to derive the Langmiur isotherm.

In particular, within the framework of Langmuir's approximation, adsorbed particles interact only with the centers of adsorption, they do not interact with each other and the energy of adsorption remains constant in all 'places' of adsorption, irrespective of the surface covering degree. Besides, the adsorption involves only a monolayer close to the surface. Generalization of these assumptions on the melts, certainly, is inconvenient, and, despite of quite good consent achieved in the ecc description for various systems by different authors [5-11], it is realized only as a starting point for the development of more exact models.

The derivation of the general dependence of the interfacial tension on the structure and temperature, published earlier [12, 13], is free of drawbacks of Langmiur's approximation. In our opinion, this approach is seldom used, and it is applied to the ecc analysis in the present work.

## 2. Thermodynamic equation for interfacial tension

Let us consider a system composed of two condensed phases *a* and *d* separated by the interfacial layers **c** and **d** (Fig. 1). The quantities corresponding the first phase, we shall denote with the superscript <'>, and those for the second phase with <">. The quantities in the bulk of the phases have a superscript $V$, and those in the interfacial layers $\omega$. The interfacial tension in such a system is equal to the work necessary to form a unit area of the new interface. We may present it in a form:

$$\sigma = \sigma' + \sigma'' \qquad (1)$$

where the interfacial tensions $\sigma'$ and $\sigma''$ differ from the surface tensions $(\sigma')_0$ and $(\sigma'')_0$ of the same non-interacting phases which have not been put into contact, by the quantities $\Delta\sigma'$ and $\Delta\sigma''$. They are due to particle interaction in the corresponding phases. The interaction manifests itself in the substance exchange, the formation of chemical bonds between particles of the phases, the redistribution of electrons in frontier layers, etc. According to [1, 12], the interfacial tension $\sigma'$ is determined by the difference of chemical potentials (cp) in a superficial layer and in the bulk of the phase:

$$\sigma' = \frac{1}{(\omega_i)'}\left[(\mu_i^\omega)' - (\mu_i)'\right] \qquad (2)$$

where the sum is taken over all components, and where $\omega_i'$ is the partial molar area, $(\mu_i)'$ is cp of $i$-th component of a phase (a particle of $i$ type), $(\mu_i^\omega)'$ is an effective quantity, the cp of $i$-type particles in a superficial layer, provided that the surface tension is removed [1, 4]. For a phase <"> similar expression is valid. Thus, the interfacial layer is formally divided into two parts **b** and **c**, each of which is put down between the volume of one of the phases and the second part of interfacial layer (Fig. 1).

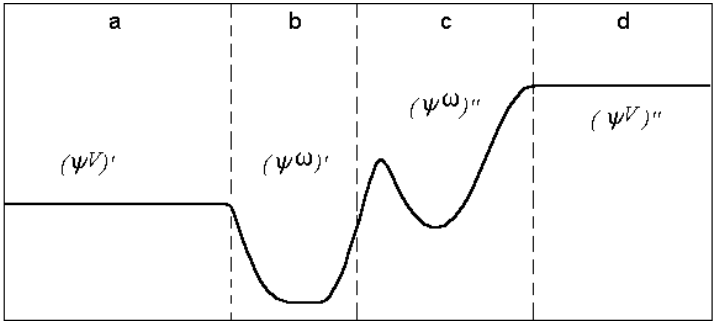

*Fig. 1. Schematic distribution of the electric potential in the vicinity of the interfacial boundary: **a** – a bulk of the metal; **b** - a superficial layer on the metal side; **c** - a superficial layer from the melt side, **d** – a bulk of the melt.*

In the following we shall restrict our consideration to liquid metal – ionic melt system in which the ions having both positive and negative electric charges will be meant as particles. In an external electric field with a potential difference $\varphi=(\psi^V)'-(\psi^V)''$ it is useful to replace cp of the particles with the electrochemical potential (ecp):

$$\bar\mu_i = \mu_i + n_i F \psi_{phase} \qquad (3)$$

where $n_i$ is the charge number of an ion, $F$ is the Faraday constant.

Popel' and Pavlov [12] have suggested to separate cp into the components, first of which corresponds to the ideal system ('*id*') of non-interacting particles, and the second one is responsible for the interaction ('*int*'). We shall assume, as it seems quite reasonable, that the term describing inter-particle interaction does not depend on the electrode potential directly:

$$\bar\mu_i = \bar\mu_{i(id)} + \mu_{i(\text{int})} \qquad (4)$$

Here one can note the analogy with the behavior of particles in the gravitational field that affects the particles' interaction in a minimal way.

Using the ideal gas thermodynamics we shall rewrite the first difference in the first phase as:

$$(\mu_{i(id)}^{\omega})' - (\mu_{i(id)})' = RT(\ln \frac{(N_i^{\omega})'}{(N_i)'} + \ln \frac{(V)'}{(V^{\omega})'}), \quad (5)$$

where $(N_i)'$ and $(N_i^{\omega})'$ are the molar fractions, and $(V)'$ and $(V^{\omega})'$ are the molar volumes in the bulk and in a superficial layer of the phase $<'>$, respectively. The expression for the second phase is similar. Thus,

$$\sigma = \frac{1}{(\omega_i)'} \left\{ RT \ln \frac{(N_i^{\omega})'(V)'}{(N)'(V^{\omega})'} + n_i F(\psi^{\omega})' - n_i F(\psi^V)' \right\} + \frac{1}{(\omega_i)''} \left\{ RT \ln \frac{(N_i^{\omega})''(V)''}{(N)''V^{\omega})''} + \right.$$
$$\left. + n_i F(\psi^{\omega})'' - n_i F(\psi^V)'' \right\} + \frac{1}{(\omega_i)'}((\mu_{i(int)}^{\omega})' - (\mu_{i(int)})') + \frac{1}{(\omega_i)''}((\mu_{i(int)}^{\omega})'' - (\mu_{i(int)})''). \quad (6)$$

The part of electrochemical potential, which is caused by the interaction in a superficial layer, was represented, as explained earlier, as an additive contribution of the interaction ecp in the bulk of the contacting phases. Contrary to [12, 13], we shall express, what is more strict, the interaction cp in a superficial layer in terms of additive contributions of those phase layers, which directly adjoin a superficial layer. For ecp of a superficial layer on each side of the phases we have:

$$\left(\mu_{i(int)}^{\omega}\right)' = \alpha_i' \left(\mu_{i(int)}^{\omega}\right)'' + (1-\alpha_i')\mu_{i(int)}';$$
$$\left(\mu_{i(int)}^{\omega}\right)'' = \alpha_i'' \left(\mu_{i(int)}^{\omega}\right)' + (1-\alpha_i'')\mu_{i(int)}''; \quad (7)$$

In the first approximation, the parameter $\alpha_i$ characterizes the average share of bonds of $i$-th type particle in each phase directed to the second phase; the rest $(1-\alpha_i)$ are directed to the first phase. Solving the set of equations (7) we relate $\mu_{i(int)}^{\omega}$ in both phases. Now it is easy to obtain:

$$(\mu_{i(int)}^{\omega})' - (\mu_{i(int)})' = \frac{\alpha_i'(1-\alpha_i'')}{1-\alpha_i' \cdot \alpha_i''} \left[\mu_{i(int)}'' - \mu_{i(int)}'\right]. \quad (8)$$

We have similar expression for the last difference in (6):

$$(\mu_{i(int)}^{\omega})'' - (\mu_{i(int)})'' = \frac{\alpha_i''(1-\alpha_i')}{1-\alpha_i' \cdot \alpha_i''} \left[\mu_{i(int)}' - \mu_{i(int)}''\right]. \quad (9)$$

According to the commonly accepted definition [3, 4], a balance condition $\bar{\mu_i'} = \bar{\mu_i''}$ is necessary for the equilibrium of contacting phases. According to (5), taking into account interaction-dependent and ideal components, we may write down:

$$\mu_{i(int)}'' - \mu_{i(int)}' = \mu_{i(id)}' - \mu_{i(id)}'' + n_i F\varphi = RT \ln\left(\frac{c_i'}{c_i''}\right) + n_i F\varphi. \quad (10)$$

In view of the above-mentioned equations it is possible to obtain expressions for the molar fractions of $i$-th component on both sides of the surface:

$$\left(N_i^\omega\right)' = N_i' \frac{(V^\omega)'}{V'} \left(\frac{c_i''}{c_i'}\right)^{\beta_i^{(1)}} \exp\left(\frac{\sigma' \omega_i' - \beta_i^{(1)} n_i F \varphi}{RT}\right);$$

$$\left(N_i^\omega\right)'' = N_i'' \frac{(V^\omega)''}{V''} \left(\frac{c_i'}{c_i''}\right)^{\beta_i^{(2)}} \exp\left(\frac{\sigma'' \omega_i'' - \beta_i^{(2)} n_i F \varphi}{RT}\right) \quad (11)$$

Here $c_i'$ and $c_i''$ are the actual concentrations of $i$-th component in the corresponding phases. We have introduced here $\beta_i^{(1)} = \frac{\alpha_i'(1-\alpha_i'')}{1-\alpha_i'\alpha_i''}$, $\beta_i^{(2)} = \frac{\alpha_i''(1-\alpha_i')}{1-\alpha_i'\alpha_i''}$.

Also it is considered that $\sum_i N_i^\omega = 1$.

Substituting (10) into equations (8) and (9), we obtain:

$$\sigma = \frac{1}{(\omega_i)'}\left\{RT \ln \frac{(N_i^\omega)'(V)'}{(N)'(V^\omega)'} + n_i F(\psi^\omega)' - n_i F(\psi^V)'\right\} + \frac{1}{(\omega_i)''}\left\{RT \ln \frac{(N_i^\omega)''(V)''}{(N)''V^\omega)''} + n_i F(\psi^\omega)'' - n_i F(\psi^V)''\right\}$$

$$+ \frac{1}{(\omega_i)'}\left[\beta_i^{(1)}\left[R\cdot T\cdot \ln(\frac{c_i'}{c_i''}) + n_i \cdot F \cdot \varphi\right]\right] + \frac{1}{(\omega_i)''}\left[\beta_i^{(2)}\left[R\cdot T\cdot \ln(\frac{c_i''}{c_i'}) - n_i \cdot F \cdot \varphi\right]\right].$$

(12)

Keeping in mind equation (1), let us write down 'the sum rule' for the surface phases separately (cf. analogous expressions in [2]):

$$\sum_i (N_i')\left(\frac{c_i''}{c_i'}\right)^{\beta_i^{(1)}} \exp\left(\frac{\sigma'\omega_i' - n_i F\cdot(\beta_i^{(1)}\varphi + (\psi^\omega)'-(\psi^V)')}{RT}\right) = \frac{V'}{(V^\omega)'};$$

$$\sum_i (N_i'')\left(\frac{c_i'}{c_i''}\right)^{\beta_i^{(2)}} \exp\left(\frac{\sigma''\omega_i'' + n_i F\cdot(\beta_i^{(2)}\varphi - (\psi^\omega)''+(\psi^V)'')}{RT}\right) = \frac{V''}{(V^\omega)''}$$

(13)

The equations (13) present a basis for the determination of the dependences $\sigma'$ and $\sigma''$ and the total value of $\sigma$ on the electrode potential $\varphi$, on the concentration of components and on other parameters. Note that the equations (13) have the most general form in the approximation used. They are not independent, because they are coupled by the parameters $\beta^{(1)}$ and $\beta^{(2)}$. If one assumes that the partial molar volumes are equal to each other and they are independent of the content, the partial surface areas also will be equal. Then we have for the interfacial tension:

$$\sigma = -\frac{RT}{(\omega)'}\left\{\ln \sum_i (N_i)'\left(\frac{c_i''}{c_i'}\right)^{\beta_i^{(1)}} \exp\left(-n_i F\left[\beta_i^{(1)}\varphi + (\psi^\omega)'-(\psi^V)'\right]\right)\right\} -$$

$$\frac{RT}{(\omega)''}\left\{\ln \sum_i (N_i)''\left(\frac{c_i'}{c_i''}\right)^{\beta_i^{(2)}} \exp\left(-n_i F\left[\beta_i^{(2)}\varphi - (\psi^\omega)''+(\psi^V)''\right]\right)\right\}$$

(14)

Note that the equation for $\sigma$ contains a great deal of parameters, and it can be applied for the description of experimental data only after some assumptions.

## 3. Potential modeling and electrocapillary equation

The electric potential $\psi^V$ has constant value inside both phases (see Fig. 1), and we may suppose that the peculiarities of the curve $\sigma(\varphi)$ should be governed by the quantity $\psi^\omega$ in corresponding phases of the interfacial layer. It is necessary to specify the structure of superficial layers and the distribution of the electric potential inside the phases. In the primitive case (see Fig. 2), when the potential changes by a jump at the phase boundary, and holds constant inside regions **b** and **c**, the dependence $\sigma(\varphi)$ will be determined by the capacitances of the del's, which are formed by adsorbed particles. The boundary in this case is equivalent to three capacitors connected in series. More complicated dependencies of $\psi$ in the transition region can be described by introducing the additional parameters.

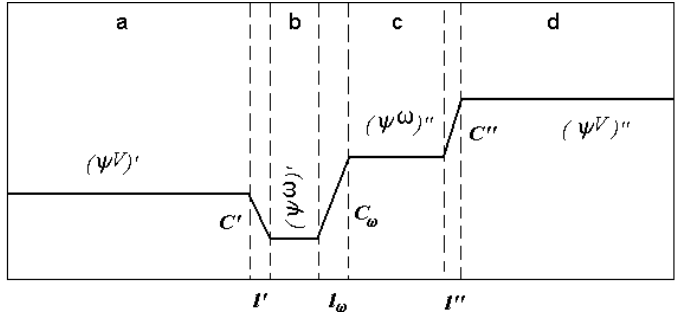

*Fig. 2. Electric potential distribution in the vicinity of interfacial boundary in the model of 'in series connected capacitors': C is the del capacitance; l is the del thickness (l<L, L is the interface region width).*

The equations (11) allow to obtain the expressions for the adsorption $\Gamma_i$ in the Guggenheim version $\Gamma_i = (N_i^\omega - N_i)/\omega$. Taking into account the adsorption of components at the interfacial boundary on both sides of phases, $\Gamma^{(n)} = \Gamma' + \Gamma''$, we have:[1]

$$\Gamma_i' = \frac{N_i'}{\omega'}\left\{\left(\frac{c_i''}{c_i'}\right)^{\beta_i^{(1)}} \exp\left(\frac{\sigma' \omega_i' - n_i F \cdot [\beta_i^{(1)}\varphi + (\psi^\omega)' - (\psi^V)']}{RT}\right) - 1\right\}. \quad (15)$$

Analogous expression is valid for the second phase.

According to [2, 12], the master equation of electrocapillarity has the form:

$$-d\sigma = (\varepsilon + \sum_i n_i F \Gamma_i) d\varphi \quad (16)$$

Here $\varepsilon$ is the charge density of one of the covers (at the electrolyte side, or at the metal one) of the del, $\Gamma_i$ is total adsorption of $i$ component, and $\sum_i \Gamma_i' n_i F = \varepsilon$, $\sum_i \Gamma_i'' n_i F = -\varepsilon$.

We assume in the following that the metal – electrolyte boundary can be treated as some transition region in which the interfacial tension is determined by two additives: by the charge of ions and electrons transferred from the bulk to the superficial layer. This charge determines the change in electrode potential. Another additive is determined by the adsorption of the components of the phases (both charged and non-charged), which chemical potential changes in the vicinity of del. The particle, in the common case, gives the input to $\sigma$ twice: at first, by the adsorption at one of the sides of the phases, and secondly, by transferring its charge to the del.

---

[1] Let's notice in passing, that a choice of the reference volume in the bulk of phases agrees with the Guggenheim [4] approach. At certain assumptions, it is possible to obtain the explicit expression for $\sigma$ as the function of electrode potential $\varphi$ and other parameters from (14), having in mind that: $\sum_i \Gamma_i' = \sum_i \Gamma_i'' = 0$.

Substituting expression (15) into (16) and integrating, we obtain:

$$\sigma' = \sigma'_o - \int_{\varphi_o}^{\varphi} \left\{ \varepsilon' - \sum_i \frac{N'_i}{\omega'} \left[ \left(c''_i/c'_i\right)^{\beta^{(1)}_i} \exp\left(\sigma'\omega' - n_i F \cdot [\beta^{(1)}_i \varphi' + (\psi^\omega)' - (\psi^V)']/RT\right) - 1 \right] \right\} d\varphi' \quad (17)$$

The expression for the quantity $\sigma''$ is similar to (17). Here $\varphi_0$ is the zero-charge potential (zcp) with the surface covering degree equals zero.[2] Substituting the potential difference in the exponential argument with the capacitance of corresponding del, and taking into account the relation of $\varepsilon'$ and integral capacity $C'_s$, namely, $\varepsilon' = C'_s(\varphi-\varphi_o)$, the following form of the equation (17) can be obtained:

$$\sigma'(\varphi) = \sigma'_o - \frac{C'_s}{2}(\varphi-\varphi_o)^2 - \int_{\varphi_o}^{\varphi} \left\{ \sum_i \frac{N'_i}{\omega'} \left[ \left(c''_i/c'_i\right)^{\beta^{(1)}_i} \exp\left(\sigma'(\varphi')\omega' - n_i F \cdot (\beta^{(1)}_i \varphi' + \varepsilon'/C')/RT\right) - 1 \right] \right\} d\varphi' \quad (18)$$

The equation (18) has the form of particular case of the Volterra equation and it gives in a closed form the electrocapillarity equation in view of Popel' – Pavlov isotherm.

Let us consider now more simple potential distribution. The transition region of the superficial layers **b** and **c** in this case can be treated as the united interfacial layer. As it was done earlier, let us assume that the electric potential holds constant inside the interfacial layer $\psi^\omega$ and it changes abruptly at the phase boundaries (see Fig. 3).

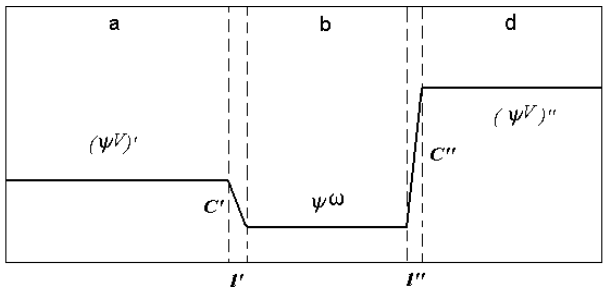

*Fig. 3. Electric potential distribution in the vicinity of the boundary in the model of two capacitors connected in series: C' and C'' are del capacitances, l' and l'' are the del's thickness, (l<L, L is the interfacial layer thickness).*

The interfacial tension can be determined in two ways according to formula (2), that means, by the differences of electrochemical potentials in the surface layer and in the bulk of one of the phases (at the metal side, or in the ionic

---

[2] For the application of the equation obtained to the analysis of ecc, it is reasonable to note that, as a rule [5−10], the content of some component at the metal – ionic melt interface which is engaged in the electrode process from one of the phases is such that its chemical (or electrochemical) potential does not depend on $\varphi$. According to the Nernst equations, this allows to pass from chemical potentials to $\varphi$ [2]:

$$d\mu_{Ox} - d\mu_{Red} = n_i F d\varphi .$$

melt). Repeating the derivation of equation (14) in this case and accounting for the fact that now there is only one average value of the surface layer parameters (region **b** in Fig. 3), one may obtain:

$$\sigma' = -\frac{RT}{\omega'}\left\{\ln\sum_i (N_i)'(c_i'/c_i'')^{\alpha_i} \exp\left(n_i F\left[\alpha_i \varphi + (\psi^{\omega}) - (\psi^V)'\right]\right)\right\} \quad (19)$$

Again, substituting the potential difference in the exponential argument with the capacitance of corresponding del, and taking into account the relation of $\varepsilon'$ and integral capacity $C'_s$, namely, $\varepsilon' = C'_s(\varphi-\varphi_o)$, we receive:

$$\sigma' = \sigma'_0 - \frac{C'_s}{2}(\varphi-\varphi_0)^2 - \int_{\varphi_0}^{\varphi}\left\{\sum_i \frac{N_i'}{\omega'}\left[(c_i'/c_i'')^{\alpha_i}\exp(n_i F\cdot(\alpha_i\varphi + \varepsilon'/C')/RT)-1\right]\right\}d\varphi \quad (20)$$

The obtained expression presents the dependence of interfacial tension on the potential, without prepositions that are characteristic of the Langmiur approach, in an explicit form, and it may be applied to the analysis of experimental ecc's, just like it has been carried out earlier in [6] in the Langmuir approximation.

The solutions of these equations can be obtained by iteration procedure, provided that the integral part does not give a great contribution. In the first approximation in this case, the integral can be taken by substituting the first two terms of the right hand sides in (18) and (20). For the determination of the total value of $\sigma(\varphi)$ the equation (18) (or (20)) must be solved for both phases.

## 4. Conclusion

The formulae proposed by Popel' and Pavlov are generalized by taking into account the effect of the electric field on the adsorption of surface-active particles. Under the conditions of constant electric potential in the surface phases, and of abrupt jumps at the superficial layers (the model of in-series connected capacitors), the integral equation is derived for the interfacial tension with the parameters having explicit physical meaning. Thus, solving these equations by iteration procedure in several versions of the models for the potential behavior in the surface area, the theoretical ecc's can be obtained for comparison of the model predictions with the experimental data.

## Acknowledgment

We are grateful to Prof. Anatoli I. Sotnikov for encouraging and helpful discussions.